\documentclass[a4paper,11pt]{article}

\usepackage{graphicx}
\usepackage{caption}
\usepackage{url}
\usepackage{color}
\usepackage{graphicx}
\usepackage[displaymath, mathlines]{lineno}
\usepackage[hidelinks]{hyperref}
\usepackage{amsmath,amssymb}
\usepackage{enumitem}
\usepackage{natbib}
\usepackage{times}

\usepackage[a4paper, total={160mm,247mm}]{geometry}
\newcommand{\pd}[2]{\dfrac{\partial #1}{\partial #2}}

\usepackage{titlesec}
\titleformat{\subsubsection}[runin]
{\normalfont\bfseries}{\thesubsubsection}{1em}{}
\begin{document}

\begin{center}
\textbf{\Large Supershear shock front contribution to the tsunami from the 2018 $\mathbf M_{\mathbf w}$ 7.5 Palu, Indonesia earthquake} \\[20pt]
\textcolor{blue}{\small Manuscript to appear in Geophysical Journal International}\\[20pt]

Faisal Amlani,$^{1}$ Harsha S. Bhat,$^{2,*}$ Wim J. F. Simons,$^{3}$ Alexandre Schubnel,$^{2}$ Christophe Vigny$^{2}$, Ares J. Rosakis$^{4}$, Joni Efendi$^5$, Ahmed E. Elbanna$^6$, Pierpaolo Dubernet$^{2}$ and Hasanuddin Z. Abidin$^{5,7}$

\end{center}
\vspace{-0.2cm}
\begin{enumerate}
\small
\it
\itemsep0em
\item{Department of Aerospace and Mechanical Engineering, University of Southern California, Los Angeles, CA, USA.}
\item{Laboratoire de G\'{e}ologie, \'{E}cole Normale Sup\'{e}rieure, CNRS-UMR 8538, PSL Research University, Paris, France.}
\item{Faculty of Aerospace Engineering, Delft University of Technology, Delft, Netherlands.}
\item{Graduate Aerospace Laboratories, California Institute of Technology, Pasadena, CA, USA.}
\item{BIG (Badan Informasi Geospasial / Geospatial Information Agency of Indonesia), Java, Indonesia.}
\item{Department of Civil and Environmental Engineering, University of Illinois at Urbana Champaign, Urbana, IL, USA.}
\item{Department of Geodesy and Geomatics Engineering, Institute of Technology Bandung, Bandung, Indonesia.}
\item[*] \textbf{Corresponding author}: \texttt{harsha.bhat@ens.fr}
\end{enumerate}

\noindent

\noindent{\bf Hazardous tsunamis are known to be generated predominantly at subduction zones. However, the 2018 $M_{\mathbf w}$ 7.5 Sulawesi (Indonesia) earthquake on a strike-slip fault generated a tsunami that devastated the city of Palu. The mechanism by which this tsunami originated from such an earthquake is being debated. Here we present near-field ground motion (GPS) data confirming that the earthquake attained supershear speed, i.e., a rupture speed greater than the shear wave speed of the host medium. We study the effect of this supershear rupture on tsunami generation by coupling the ground motion to a one-dimensional nonlinear shallow-water wave model accounting for both time-dependent bathymetric displacement and velocity. With the local bathymetric profile of Palu bay around a tidal station, our simulations reproduce the tsunami arrival and motions observed by CCTV cameras. We conclude that Mach (shock) fronts, generated by the supershear speed, interacted with the bathymetry and contributed to the tsunami.}

%%%%%%%%%%%%%%%%%%%%%%%%%%%%%%%%%%%%%%%%%%%%%%%%%%%%%%%%%%%%%%%%%%%%%%
% MAIN TEXT
%%%%%%%%%%%%%%%%%%%%%%%%%%%%%%%%%%%%%%%%%%%%%%%%%%%%%%%%%%%%%%%%%%%%%%
\vspace{0.2cm}
\noindent

\section{Introduction}

Tsunamis are well-known to be amongst the most destructive consequences of earthquakes \citep{bryant2008tsunami,synolakis2005,pugh2014, robke2017}, and the 2018 Sulawesi earthquake was no exception: it generated a devastating tsunami \citep{mai2019,fritz2018} in the nearby Palu bay in which hundreds were killed and tens of thousands more displaced from their homes \citep{asean2018}. However, this was a very unexpected occurence since the earthquake was associated with predominantly in-plane ground motion produced by strike-slip ruptures. 

These motions are not known to excite significant tsunamis, the underlying physical mechanisms behind the tsunami have largely remained a mystery \citep{syamsidik2019}. 

Studies conducted to explain the phenomenon have not arrived at definitive conclusions \citep{muhari2018} nor have adequately captured observed records \citep{jamelot2019,heidarzadeh2019,ulrich2019}; the main consensus appears to be that some form of ground motion (e.g., landslides \citep{sassa2019} or the reverse-slip motion of the fault \citep{he2019}), amplified by the bay, is to blame.

However, a key notable feature of this earthquake is that it ruptured at supershear speed \citep{bao2019,socquet2019}, which results in a manifestation of shear and Rayleigh Mach fronts carrying substantial vertical velocity with relatively slow attenuation over large distances \citep{bernard2005,dunham2008a}. The existence of supershear earthquakes has been proven theoretically and experimentally since the early 1970s \citep{burridge1973,andrews1976,das1977, wu1972,rosakis1999a,xia2004,passelegue2013}. The 1979 $M_{w}$~6.5 Imperial Valley (California) earthquake was the first naturally observed supershear earthquake rupture \citep{archuleta1984}. Since then, several more (although rare) earthquakes have been recorded to propagate at supershear speeds: the $M_{w}$~7.4 1999 Izmit in Turkey \citep{bouchon2001}, the $M_{w}$~7.8 2001 Kunlun \citep{robinson2006a} and the $M_{w}$~7.8 2002 Denali \citep{ellsworth2004b,mello2014}, to name a few.

Although the overall tsunami behaviour at Palu is likely a combination of several effects that include these supershear dynamics as well as landslides, recent studies \citep{oral2020,ulrich2019, jamelot2019} suggest that the influence from phenomena such as the latter may be secondary: the rupture itself may have created adequate seafloor movement to excite the tsunami, which was subsequently amplified by the shallow and narrow two-dimensional (2D)/3D geometric features of the Palu bay. {Indeed, high-frequency waveform observations (1Hz) from carefully calibrated analysis of CCTV and social media camera footage near the Pantoloan (PANT) station suggest a near instantaneous, high-frequency, tsunami arrival \citep{carvajal2019}---consistent with a coseismic source near the coast. This arrival is not captured by observations that were made by the one working acoustic sensor at the PANT tidal gauge \citep{sepulveda2020}, whose resolution (0.02Hz, or one measurement per minute) is too coarse to have captured the much shorter wavelength (a 1-2 minute period \citep{carvajal2019}) of the tsunami as observed by the high-resolution camera analysis.} 

{Hence the primary objective of this work is to explain the near instantaneous arrival of the tsunami by elucidating the tsunami generation process of the supershear strike-slip Palu earthquake in order to more fully understand the role played by the corresponding rupture dynamics on the observed timing and first motions of the subsequent tsunami}. In particular, we incorporate a feature neglected in {previous modeling studies on Palu \citep{ulrich2019,jamelot2019}} that is a defining characteristic of supershear earthquakes: the \emph{velocity} of the ground motion \citep{bernard2005,dunham2008a}. Using a model validated by the first near-field evidence (presented later in the paper) of supershear at Palu, our results imply that ground velocities, which better represent the intricacies of the Mach fronts, may further explain the observed motions of the tsunami. Since other studies (including those investigating landslides and liquefaction) have adequately captured much of the observed run-up amplitudes and some local inundations, the scope of this paper is to focus on the arrival, first motions and phases {inferred from CCTV camera records near the PANT station \citep{carvajal2019,sepulveda2020}.}

This manuscript is organized as follows. Section 2 describes the overall methods and data employed in this study, including earthquake displacements/velocities simulated by a supershear rupture model (Section 2.1), a corresponding tsunami model (Section 2.2) that accounts for such dynamic displacements/velocities (numerically simulated via a novel  pseudo-spectral methodology for solving the shallow water wave equations), and GPS ground displacement data recorded at the PALP station during the Sulawesi earthquake (Section 2.3). The results and discussion of Section 3 provide the aforementioned evidence of supershear observed directly from those GPS records (Section 3.1), where the corresponding rupture dynamics are then numerically modeled and subsequently incorporated into the tsunami equations for comparison with observed waveforms acquired from the PANT observations (Section 3.2). Concluding remarks are provided in Section 4.

\section{Methods \& data}

%
% \subsection{\normalsize Modeling the effect of supershear velocity on tsunami generation}
%
% \subsubsection{Supershear Earthquake Dynamics and Rupture Modeling}

\subsection{Supershear modeling}

For the considered supershear earthquake dynamics and the corresponding rupture modeling (mutually validated by GPS data in Section 3.1 and subsequently employed to source the Palu tsunami configuration in Section 3.2), we use existing numerical simulations conducted by Dunham and Bhat~\citep{dunham2008a}. Such simulations have been produced by a staggered-grid finite-difference (FD) code \citep{favreau2002} with the fault boundary conditions implemented using a staggered-grid split-node (SGSN) method \citep{dalguer2007}. Since~\citet{dunham2008a} have provided non-dimensionalised solutions, we simply dimensionalise their results for the Palu earthquake by using a shear modulus of $30$ GPa, stress drop of $20$ MPa and a shear wave speed of $3.5$ km/s. The depth of the rupture is assumed to be $7.5$ km. These parameters, reasonable for crustal earthquakes, were chosen to best fit the observations. The resulting particle velocities and displacements are presented in Section 3.

% in Fig.~2.

\subsection{Tsunami modeling}

\subsubsection{Governing shallow water wave equations with dynamic ground displacement \& velocity}

Using the synthetic particle motions generated by the 3D supershear earthquake model described above (which, as later discussed in Section 3.1, agree with PALP GPS records {and are reasonably assumed to sweep past the bay near Pantoloan}), a 1D non-linear shallow water wave model incorporating time-dependent ground movements of velocity and displacement \citep{dutykh2016} is utilised to simulate the generation and propagation of the tsunami. Such a model employs the depth-averaged shallow water approximation of the Euler equations, which can be written as a system of coupled hyperbolic partial differential equations given by

\begin{linenomath*}
\begin{equation}\label{eq:swe}
\begin{cases}
\pd{H}{t} + \pd{(Hu)}{y} = 0,\\
\pd{(Hu)}{t}+\pd{(Hu^2)}{y}+gH\pd{\eta}{y} =0,
\end{cases} \quad 0\leq y \leq L, \quad t\geq 0.
\end{equation}
\end{linenomath*}

Here, $u(y,t)$ is the fluid velocity, $\eta(y,t)$ is the sea surface height and $H(y,t) = \eta(y,t)+ h_0(y){-h(y,t)}$ is the absolute height from the bed-level to the water surface for an initial at-rest bathymetry $h_0(y)$. The constant $g$ is the acceleration due to gravity. The entire domain of length $L$ is subjected to a time-dependent ground perturbation $h(y,t)$ which---together with the corresponding ground velocity $\partial h(y,t)/\partial t$ included in Equation~\eqref{eq:swe}---sources the subsequent tsunami dynamics. In the specific Palu bay configuration considered in this work (Section 3.2), these values are determined from the 3D supershear earthquake model as discussed in Section 3.1.

\subsubsection{Pseudo-spectral numerical analysis based on Fourier continuation}

The complete non-linear system given by~\eqref{eq:swe} is solved using a numerical scheme based on an accelerated Fourier continuation (FC) methodology for accurate Fourier expansions of non-periodic functions \citep{lyonbrunoII,albinbruno,amlanibruno}. Considering an equispaced Cartesian spatial grid on, for example, the unit interval $[0,1]$ (given by the discrete points $y_i = i/(N-1), i=0,\dots,N-1$), Fourier continuation algorithms append a small number of points to the discretised function values $\eta(y_i), u(y_i)$ in order to form ($1+d$)-periodic trigonometric polynomials $\eta_\text{cont}(y), u_\text{cont}(y)$ that are of the form

\begin{equation}\label{eq:fcseries}
\eta_{\text{cont}}(y) = \displaystyle\sum_{k=-M}^{M} a_k e^{\frac{2\pi i k y}{1+d}} ~~, ~~ u_{\text{cont}}(y) = \displaystyle\sum_{k=-M}^{M} b_k e^{\frac{2\pi i k y}{1+d}}
\end{equation}

and that match the given discrete values of $\eta(y_i), u(y_i)$, i.e., $\eta_{\text{cont}}(y_i) = \eta(y_i), u_{\text{cont}}(y_i) = u(y_i)$ for $i=0,...,N-1.$ Spatial derivatives for the shallow water system are then computed by exact term-wise differentiation of~\eqref{eq:fcseries} as

\begin{equation}\label{eq:termwise}
\begin{aligned}
\pd{\eta}{y}(y_i) & = \pd{\eta_\text{cont}}{y}(y_i) & = \displaystyle\sum_{k=-M}^{M} \left(\frac{2\pi i k}{1+d}\right) a_k e^{\frac{2\pi i k y_i}{1+d}},\\
\pd{u}{y}(y_i) & = \pd{u_\text{cont}}{y}(y_i) & = \displaystyle\sum_{k=-M}^{M} \left(\frac{2\pi i k}{1+d}\right) b_k e^{\frac{2\pi i k y_i}{L+d}}.
\end{aligned}
\end{equation}

In essence, FC algorithms add a (fixed) handful of additional values to the original discretized function in order to form a  periodic extension in $[1,1+d]$ that transitions smoothly
from $\eta(1)$ back to $\eta(0)$ (similarly for $u$). The resulting continued functions can be viewed as sets of discrete values of periodic and smooth functions that can be approximated to high-order on slightly larger intervals by a trigonometric polynomial. Once these discrete periodic continuation functions have been constructed, corresponding Fourier coefficients $a_k, b_k$ in Equation~\eqref{eq:fcseries} can be obtained rapidly from applications of the Fast Fourier Transform (FFT). The adopted FC parameters employed in this work as well as a detailed presentation on the accelerated construction of FC functions can be found in~\citet{amlanibruno}.

Employing these discrete continuations in order to evaluate spatial function values and derivatives on the discretised physical domain modeled by the shallow water wave equations, the algorithm is completed by employing the explicit fourth-order Adams-Bashforth scheme \citep{amlanibruno,amlaniniema,amlaniwei} to integrate the corresponding ordinary differential equations in time from the given initial conditions $\eta(y_i,t)= u(y_i,t) = 0$ up to a final given time. The final full solver enables high-order accuracy and nearly dispersionless resolution of propagating waves with mild, linear Courant-Friedrichs-Lewy constraints on the temporal discretisation---properties that are important for adequate resolution of the different spatial and temporal scales involved between the supershear source dynamics and the subsequent tsunami dynamics.  Both implicit and explicit FC-based partial differential equation solvers have been successfully constructed and utilised for a variety of physical problems including those governed by radiative transfer equations \citep{gaggiolibruno}, classical wave and diffusion equations \citep{lyonbrunoII,brunoprieto}, Euler equations \citep{shz}, convection-diffusion equations~\citep{amlaniwei}, Navier-Cauchy elastodynamics equations \citep{amlanibruno,amlanicarlos}, Navier-Stokes fluid equations \citep{albinbruno, cubillos,fontanabruno}, and fluid-structure hemodynamics equations \citep{amlaniniema}.

\subsection{GPS data from the Palu earthquake}
The dual-frequency GPS is processed using the scientific GIPSY-OASIS II software version 6.4 \citep{webb1997}. The (post-processing) Precise Point Positioning (PPP) method \citep{zumberge1997} is used in kinematic ($1$ s) mode to derive precise absolute coordinates for the PALP station. Precise ephemeris of GPS satellites (non-fiducial style, using high-rate $30$ s clocks) along with Earth rotation parameters (ERP) in the IGS14 framework \citep{rebischung2016} are obtained from the Jet Propulsion Laboratory (JPL). A satellite elevation mask angle of $7$ degrees and absolute International GNSS Service (IGS) antenna phase centre corrections are applied. The Vienna tropospheric Mapping Functions (VMF1) are used in estimating both zenith delay and gradients, downloaded from the Global Geodetic Observing System website (\url{http://vmf.geo.tuwien.ac.at/}). The global ocean tide model applied in the GPS data processing is (FES2014b), and the ocean loading parameters have been retrieved from the Onsala Space Observatory website (\url{http://holt.oso.chalmers.se/loading/}). To enhance the coordinate solutions, the daily global wide lane phase bias (wlpb) files from JPL have been used to resolve the phase cycle ambiguities \citep{bertiger2010}. Although each kinematic position has a higher uncertainty and is affected by biases which usually cancel out over long periods of measurements, the instantaneous co-seismic displacements at PALP are much higher than the high-frequency noise of around $1$ cm and $2$ to $3$ cm for, respectively, the horizontal and vertical positions. Finally, the GPS time tags are corrected to UTC time by subtracting $18$ s. The co-seismic displacement of the station simply follows from epoch-to-epoch coordinate differences. The standard available script has been modified to properly weigh the phase/code measurements of the stations and also to output the correlations. The X-Y-Z Cartesian component positions are then converted to the north-east-up positions along with their formal standard deviations. They are scaled using the weighted-root-mean-square of all the positions up to the time of the earthquake and generally reach a relative precision ($3\sigma$) of about $30$ mm on the horizontal components. The resulting displacement field is then differentiated by computing adaptive linear fits adapted to satisfy an error to fit criteria. The slope of the linear fit then gives the local velocity. The resulting data is then resampled again at 1Hz by linear interpolation. The corresponding velocity data is presented in Section 3.

\section{Results \& discussion}

\subsection{\normalsize Direct evidence of a supershear rupture}

In this section we provide the first-ever observation of a supershear rupture by a high-rate GPS station. 

\begin{figure*}%[tbh!]
\centering
\includegraphics[width=\textwidth]{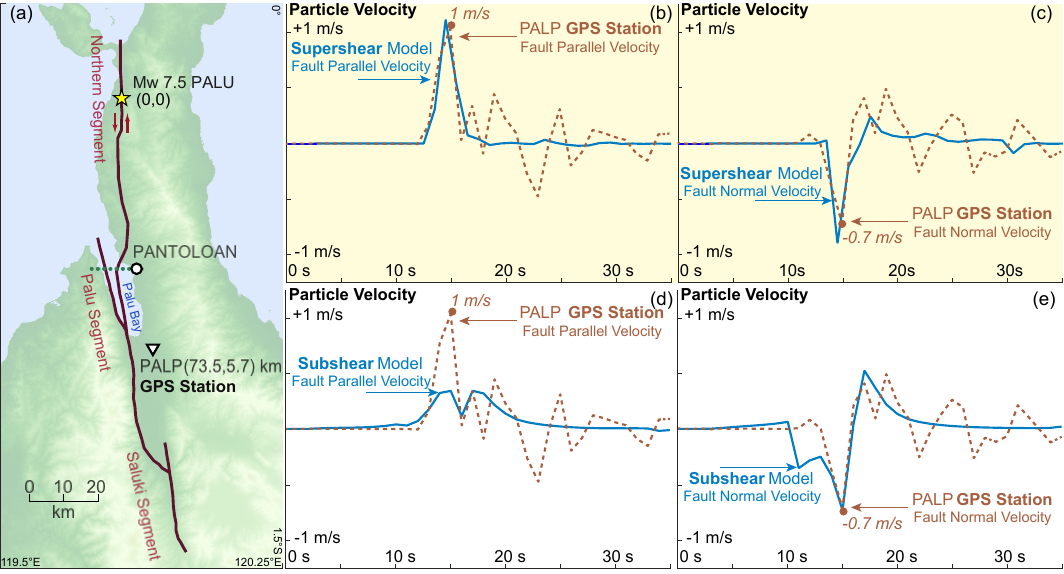}
\caption{{The earthquake rupture and near-field evidence of supershear.} (a) The Palu-Koro fault system, where the Pantoloan tidal gauge and the PALP GPS station are marked. The green line of dots represents the slice of the bay considered for the tsunami model employed in this study. (b) Comparison between the fault parallel particle velocities recorded at the PALP station with those generated by the numerical supershear rupture model \citep{dunham2008a}. (c) Comparison between the corresponding fault normal particle velocities. (d,e) Same as (b,c) but for a subshear rupture.}
\label{fig:fig1}
\end{figure*}

The most unmistakable signature of a supershear rupture is that the fault parallel particle velocity dominates over the fault normal velocity \citep{dunham2005,mello2014} (when the rupture velocity $v$ is greater than $\sqrt{2}c_s$ for a shear wave speed $c_s$). The opposite signature is expected for a  subshear rupture. Figure~\ref{fig:fig1}a shows the Palu-Koro fault system (comprising of the three segments) with the location of the high-rate, 1Hz, PALP GPS station. Figures~\ref{fig:fig1}b-c show the particle velocities recorded during the Sulawesi earthquake, clearly demonstrating a fault parallel particle velocity greater than the fault normal velocity ($\sim\!\!1.0$ m/s versus $\sim\!\!0.7$ m/s). This proves that the rupture, as it passed by the PALP station, definitively went supershear and hence attained a speed between $\sqrt{2}c_s$ and the P-wave speed,  $c_p$, of the medium (the absolute limiting speed of the rupture). This represents the first-ever observation of a supershear rupture by a high-rate GPS station. \citet{socquet2019} and \citet{bao2019} have also inferred that this earthquake went supershear, but mainly through far-field observations employing geodetic and teleseismic data, respectively. The only other near-field evidence of a supershear earthquake was obtained using an accelerometer (250Hz) at Pump Station 10 (PS10) during the 2002 Mw 7.9 Denali earthquake \citep{ellsworth2004b,mello2014}. We emphasize here that we have not performed any kinematic inversion of the GPS data; we instead have employed well-known unique signatures of near source ground velocity for supershear ruptures~\citep{dunham2005,mello2014} that indubitably confirm that the rupture, at least as it passed by the PALP station, was supershear.

We can further compare the PALP records against a 3D supershear earthquake simulation (Section 2.1) whose rupture propagates at a speed of $v = 1.6c_s$ and whose corresponding particle velocities are computed at 100Hz and then decimated to match the 1Hz sampling rate of the GPS observations.  The synthetic data and the GPS records are in excellent agreement for the main rupture pulse (Figures~\ref{fig:fig1}b-c). Subsequent arrivals are not as well-captured since the numerical model does not account for local velocity structure nor detailed fault geometry. A similar comparison with synthetic velocities computed for a subshear rupture ($v = 0.8c_s$) finds that they are in poor agreement with GPS data (Figures~\ref{fig:fig1}d-e). This clearly suggests that the supershear rupture speed was $1.6c_s$ (around $5.3$ km/s) when it passed by PALP (\citet{ulrich2019} also find a speed greater than $\sqrt{2} c_s$). We have thus provided the definitive first near-field high-rate GPS-based proof that the Sulawesi earthquake rupture actually did go supershear as claimed and, further, have validated the numerical data employed to source the tsunami model in what follows.

\subsection{\normalsize Capturing the arrival and first motions at Pantoloan}

\begin{figure*}
\centering
\includegraphics[width=\textwidth]{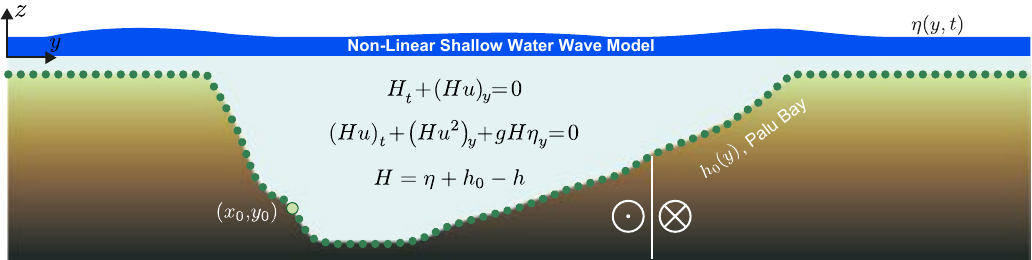}
\caption{Diagram of the non-linear shallow water wave system for tsunami height $\eta$, initial bathymetry $h_0$ (basin width 9.2 km, maximum depth 710 m) and bathymetry perturbation (source) $h$.}
\label{fig:fig2}
\end{figure*}

The specific Palu bay configuration is outlined in Figure~\ref{fig:fig2} along the horizontal $y$-axis, where $z = \eta(y,t)$ represents the water height relative to the background sea level. The bathymetry shape closely approximates that of the segment demarcated by the dotted green line near the Pantoloan tidal gauge in Figure~\ref{fig:fig1}a (basin width 9.2 km, maximum depth 710 m and an average slope of $7^{\circ}$ to the east and $27^{\circ}$ to the west of the bay \citep{weatherall2015}). The shallowest part is taken to be $1$ m, and the distance between the virtual gauge and the fault is $4.3$ km. The complete computational domain is taken to be twice the basin width ($L=18.4$ km). 

\begin{figure}
\centering
\includegraphics[width=0.65\textwidth]{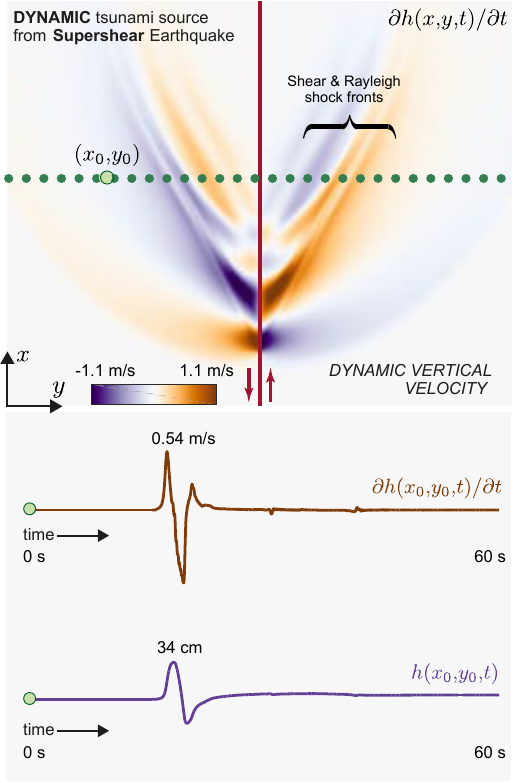}
\caption{Snapshot of the dynamic vertical velocity from a supershear earthquake with its temporal evolution at an example point $(x_0,y_0)$ (light green circle). The dark green dots correspond to the source locations used to perturb the bathymetry domain in Figure~\ref{fig:fig2}}
\label{fig:fig3}
\end{figure}

\begin{figure}
\centering
\includegraphics[width=0.65\textwidth]{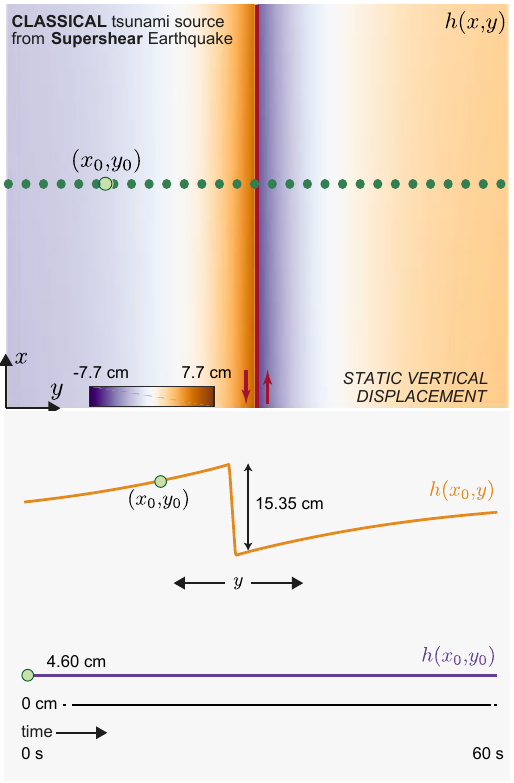}
\caption{The static displacement field due to a supershear earthquake and the spatial profile in $y$ of the static displacement field with its temporal evolution at an example point $(x_0,y_0)$ (light green circle).  The dark green dots correspond to the source locations used to perturb the bathymetry domain in Figure~\ref{fig:fig2}.}
\label{fig:fig4}
\end{figure}

Figure~\ref{fig:fig3} presents a temporal snapshot in the $(x,y)$-plane (the ground surface) illustrating the dynamic vertical velocity field (and associated Mach fronts) which is input as a synthetic source in conjunction with its corresponding time-dependent displacement field. The fault and the sense of slip (left-lateral) are indicated in red, and the data applied to perturb the bathymetry is taken along the green dotted line (whose locations correspond to the same markers indicated in Figure~\ref{fig:fig2}). For an example point located at $(x_0,y_0)$ and highlighted in a larger light green circle, Figure~\ref{fig:fig3} additionally presents the temporal evolution of both the vertical velocity (which can reach $\sim\!\!1$ m/s along the domain) as well as its corresponding ground displacement (which, in the 1D setting, can reach $\sim\!\!40$ cm). As already noted, the shapes and the maximum values of these profiles remain fairly unattenuated at large distances from the original earthquake---a hallmark of the energy carried by supershear shock fronts \citep{bernard2005,dunham2008a}. 

\begin{figure}
\centering
\includegraphics[width=\textwidth]{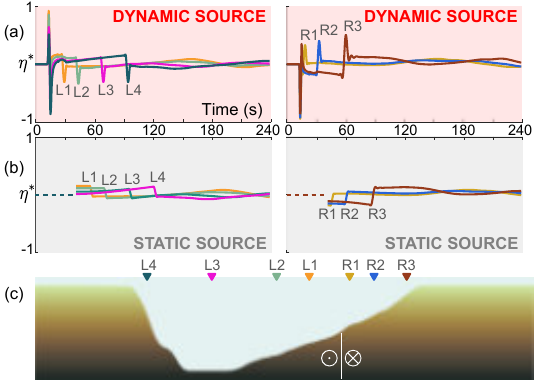}
\caption{{{Simulated tsunamis generated by dynamic and static (classical) sources.} (a,b) The time histories (sampled at 10Hz) of normalized water heights $z=\eta^{*}$ predicted at various synthetic stations (marked L1-L4 and R1-R3 in (c)) along the Palu bay for tsunamis generated by a supershear earthquake due to (a) dynamic and (b) static sources. The dashed line in (b) corresponds to the duration of the earthquake. (c) The computational domain overlaid with the locations of the synthetic stations L1-L4 and R1-R3.}}
\label{fig:fig5}
\end{figure}

For the results that follow, Figure~\ref{fig:fig4} additionally present the analogous inputs for classical modeling of seismogenic tsunamis. In a classical setting \citep{pedlosky2013}, the source is often modeled as a static displacement perturbation applied to the bathymetry (rather than dynamic ground motion), i.e., a static $h(y,t) = h(y)$ that neither accounts for the time-dependence nor the velocity of the sea floor (other simple approximations to more complicated sources are also standard \citep{tanioka1996,kajiura1963}). From the supershear earthquake results, this corresponds to the final, permanent ground displacement at the end of the profiles in Figure~\ref{fig:fig2}c and is expectedly on the order of a few centimeters.

\begin{figure}
\centering
\includegraphics[width=\textwidth]{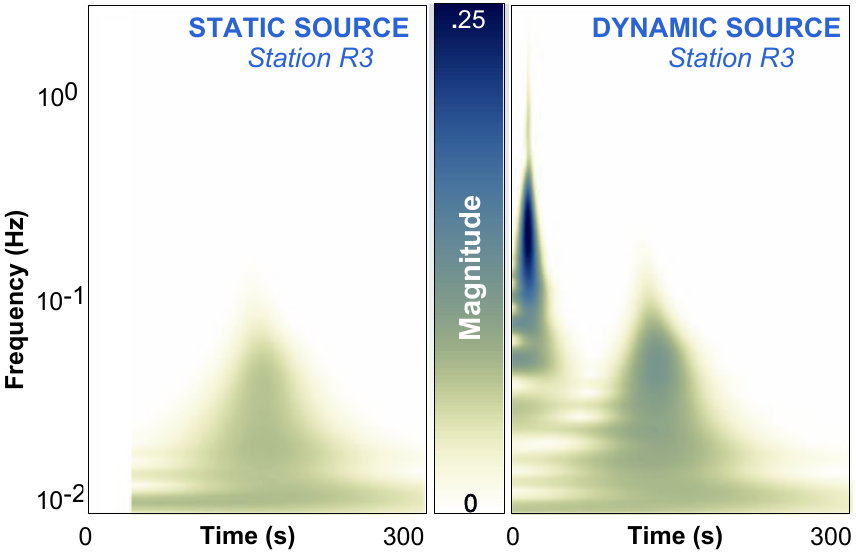}
\caption{Magnitude scalogram of the spectral contributions at the synthetic tidal gauge station R3 due to static and dynamic sources.}
\label{fig:fig6}
\end{figure}

Using such inputs with the FC-based tsunami model described in Section 2.2,  {Figures~\ref{fig:fig5}a-b} present the corresponding results of the simulated water height $z = \eta^{*}(y,t)$, normalized {by the absolute maximum from the dynamic case (i.e., $\eta^{*}(y,t) = \eta(y,t)/\text{max}_{t}|\eta_{\text{dynamic}}(y,t)|$)},  {at various synthetic stations (whose locations are indicated in Figure~\ref{fig:fig5}c)} simulated by both the dynamic and static (classical) sources generated from the same supershear earthquake simulation. Figure~\ref{fig:fig7} additionally presents the complete spatiotemporal evolution. The numerical modeling has been conducted at a much higher temporal resolution (a timestep of $\Delta t = 2.62\times10^{-3}$ s) but plotted at 10Hz. {The effects of the dynamic source, which is on the order of seconds, clearly produces high-frequency and high-amplitude waves in contrast with the static source (see Figure~\ref{fig:fig6} for a comparison of the spectral content between the two). These high-energy waves are generated earlier than those of the static case but start shedding their high energy content as they slow down in their progress towards the coastline; the two begin to resemble one another in shape (Figure~\ref{fig:fig8} presents an alternate visualisation in the form of snapshots in time across the bay). We note that, for comparison throughout, we have presented normalized water heights: since more energy of the Mach fronts is carried \emph{along} the fault \citep{bernard2005,dunham2008a} running in the direction $x$ (Figure~\ref{fig:fig2}), the 1D model in $y$ will naturally generate lower amplitudes (on the order of half a meter). However, similar tsunami signatures can still be expected and, indeed, {\citet{elbanna2020} have demonstrated that, by incorporating horizontal motions in generic 2D/3D bay-like bathymetry, similar behaviour can be observed but with amplitudes on the order of metres.}

\begin{figure}
\centering
\includegraphics[width=\textwidth]{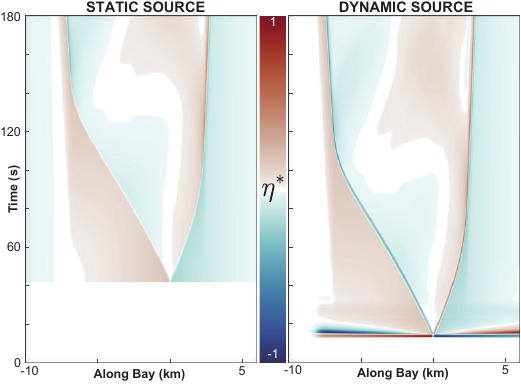}
\caption{{The complete solution, for the first five minutes, of the normalised water height $\eta^{*}$ due to (left) a static source and (right) a dynamic source (both generated from the same supershear earthquake).}}
\label{fig:fig7}
\end{figure}

{Although the final waveforms are similar, a notable feature of Figure~\ref{fig:fig5} is the earlier arrival at the coastline for the dynamic case. This is more clearly illustrated in Figure~\ref{fig:fig9}a, which presents the corresponding simulated time histories at the PANT station} (whose geographic location is indicated in Figure~\ref{fig:fig1}a) and, more importantly, presents a comparison between the waveforms of these models {with those generated at 1Hz by carefully calibrated, and timed, CCTV and other video sources in the vicinity of the PANT tidal gauge \citep{carvajal2019}. The simulations and camera records indicate an arrival that is expectedly \citep{sepulveda2020} not captured by the coarse (0.02Hz) tidal gauge at PANT, whose corresponding comparison to simulations is presented in Figure~\ref{fig:fig9}b.} Remarkably, the arrival and first motions observed from the camera records in Figure~\ref{fig:fig9}a are in excellent agreement with the 1D approximation generated by excitation from the dynamic source. Later phases, which can be attributed to wave reflections within the bay, are not as well-captured since our model does not fully account for the localized effects of the 2D/3D bathymetric profile. Nevertheless, the tsunami arrival and primary dynamics are correctly reproduced.

\begin{figure}
\centering
\includegraphics[width=0.9\textwidth]{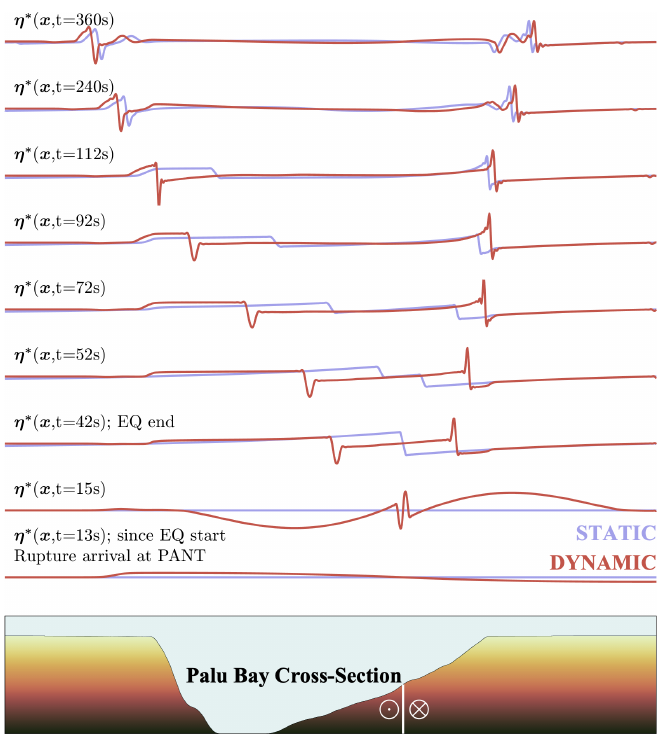}
\caption{{Simulated snapshots of normalised tsunami waves at various times generated by static source and dynamic source models along the entire Palu bay. Here, $\eta^{*}$ is the tsunami height normalised by the absolute maximum from the dynamic case, i.e., $\eta^{*}(y,t) = \eta(y,t)/\text{max}_{t}|\eta_{\text{dynamic}}(y,t)|$.}}
\label{fig:fig8}
\end{figure}

{By contrast, the static source model predicts a much later arrival. However, this is to be expected since we have employed a common approach of the classical modeling community where static results are shifted by the earthquake duration (i.e., the time taken to establish the final vertical displacement that is used for the static source, or about 42 seconds at Sulawesi \citep{usgs2018}). Some static models do account for the finite duration of the rupture by gradually increasing the static offset over this duration, but this doesn't account for the full wave dynamics of the source~\citep{satake2013}. This is a reasonable assumption for far-field tsunamis, but it is not clear that this is justifiable for a near-field source like at Palu bay, nor is it clear how much of a shift should be introduced \citep{lotto2017}. Indeed, for a fair comparison between static and dynamic models, one should wait until the end of the rupture to obtain the final static offset because of secondary slip pulses and various reflections from the surrounding medium. This can be seen in the GPS records in Figure~\ref{fig:fig1}, where non-negligible ground motion is still being recorded after the passage of the main rupture pulse. We also note that the correct timing prediction is only possible through simulations informed by the full supershear dynamics (which need not make any such assumptions), where the corresponding comparisons in Figure~\ref{fig:fig9} suggest an essentially Occam's razor explanation for the arrival observed by the PANT video waveforms: when the Sulawesi rupture went supershear, the high-frequency ground velocities carried by the shock fronts initiated a tsunami in Palu bay at the instance when the rupture swept past the station at $t\approx13$s (see also Figures~\ref{fig:fig7} and~\ref{fig:fig8}).}

\begin{figure}
\centering
\includegraphics[width=\textwidth]{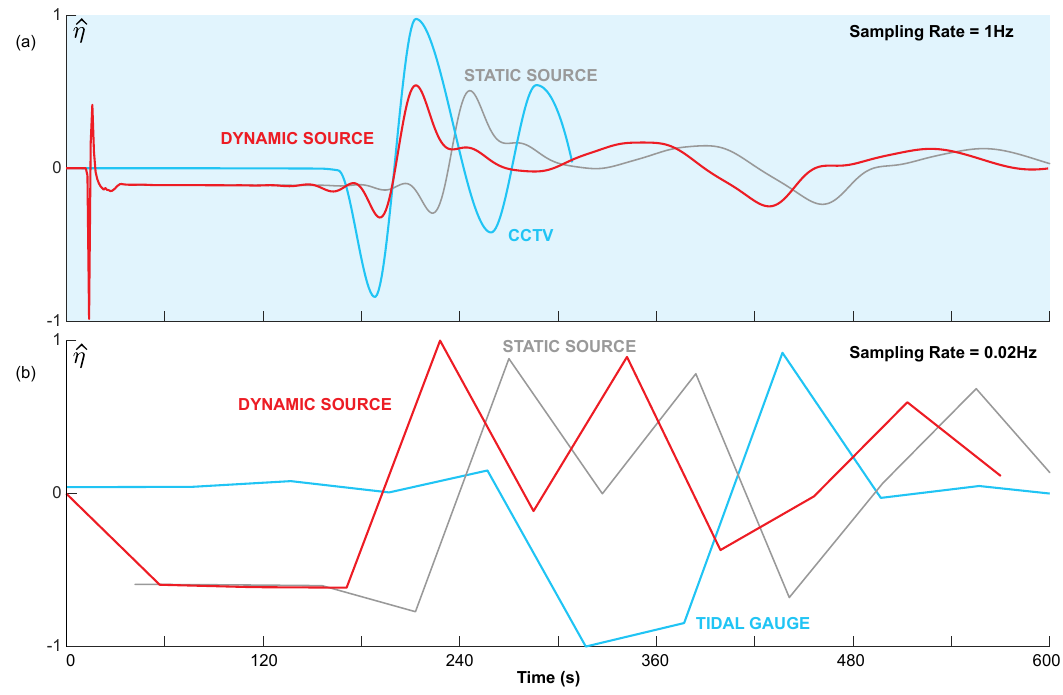}
\caption{{{Comparisons of model predictions and observations at the PANT station.} (a) The time histories of normalized water heights predicted by simulations and those observed by the high-resolution (1Hz) PANT video record waveforms obtained from the author data provided in \citep{carvajal2019}. Here, $\hat{\eta} \equiv \eta^{*} = \eta/\text{max}_{t}|\eta_{\text{dynamic}}|$ for the simulations and $\hat{\eta} = \eta_{\text{CCTV}}/\text{max}_{t}|\eta_{\text{CCTV}}|$ for the CCTV video-generated waveforms. (b) Corresponding normalized comparison with the low-resolution (0.02Hz) tidal gauge.}}
\label{fig:fig9}
\end{figure}

\section{Conclusions}
Hence we confirm that the Palu, Indonesia earthquake went supershear (via the first direct observation of such a rupture in a GPS station, accomplished here using the unique near-source signatures of supershear~\citep{dunham2005,mello2014}) and conclude that, by modeling the effects of supershear on the generation of tsunamis in a shallow geometry, the corresponding ground motion resulting from the associated Mach fronts (which carry minimally attenuated velocities to large distances) may well have contributed to the initiation of the Palu tsunami. This work provides a robust proof-of-concept, albeit in 1D, on the contribution of shock fronts in tsunami generation. In order to gain further insight into this process, more detailed modeling in 3D is needed to account for, e.g., geometrical spreading, attenuation and detailed 3D velocity structure from tomographic studies. Regardless, since nothing geologically specific about the bay, except its geometry, has been introduced, our results signify the importance of such configurations for tsunami hazard assessment due to strike-slip earthquakes. The same physical ingredients (supershear rupture and a shallow bay) may combine to produce similar effects elsewhere. Below we simply tabulate well known strike-slip faults that cut through various gulfs and bays \citep{robinson2010}.

\begin{itemize}[leftmargin=30pt]
\item Tomales bay in California which is crossed offshore by the San Andreas fault system \citep{johnson2019}
\item Izmit bay in Turkey which is crossed by the North Anatolian fault \citep{altinok2001}
\item Gulf of Tonkin in Vietnam which is intersected by the Red river fault system \citep{tapponnier1986}
\item Gulf of Martaban in Burma which is cut by the Sagaing fault \citep{vigny2003}
\item The gulf of Aqaba in the northern tip of the Read sea is crossed by the Dead Sea fault system \citep{ambraseys1994}
\item Several bays and straits in the Philippines that are cut through by the Luzon fault system \citep{yumul2003}
\end{itemize}

Some of these regions, as well as the Palu bay, have suffered from historical tsunamis. On the contrary, the 2012 off Northern Sumatra earthquake and the 2013 Craig, Alaska earthquake both went supershear but caused negligible (or no) tsunamis since they occurred in deep ocean without any shallow bay near them. Additionally, the 1999 Izmit earthquake was subshear as it passed through the Izmit bay and thus generated only a negligible tsunami. Hence we reemphasize that both the supershear rupture {\it and} a shallow bay are key to generate contributions to tsunami motions. We thus suggest that any rapid assessment of tsunami hazard after a strike-slip earthquake should also involve a rapid assessment of the earthquake rupture velocity as we have shown that ultimately the focal mechanism, the depth \textit{and the speed of the rupture} all contribute towards the generation of tsunamis.

\section*{Acknowledgements}
F.A. would like to thank Prof. N.M. Pahlevan at USC for encouraging this work. H.S.B. acknowledges the European Research Council grant PERSISMO (grant 865411) for partial support of this work. A.S. would like to acknowledge the European Research Council grant REALISM (2016-grant 681346). A.J.R. was supported by the Caltech/MCE Big Ideas Fund (BIF). A.E. acknowledges support by the National Science Foundation (CAREER Award Number 1753249). The continued (long-term) operation of the GPS stations in Central Sulawesi, Indonesia data has been co-facilitated by the EU-ASEAN SEAMERGES (2004-2006) and GEO2TECDI-1/2 projects (2009-2013) in cooperation with the Geospatial Information Agency of Indonesia (BIG). GPS data acquisition and research was partly funded by grants from the Dutch NWO User Support Programme Space Research (2007-2018). We would like to express our special thanks to the local staff of the Indonesian Meteorology, Climatology, and Geophysical Agency (BMKG) offices in Palu for hosting the GPS station and for being available 24/7 to assist us with optimal operation of the equipment. A special thanks to our local survey staff, B.R. Umar and A. Urif, for their continued support (including directly after the earthquake) in inspecting the GPS stations.

\section*{Data Availablility}
All codes are available upon reasonable request to the corresponding author. All relevant data relating to the tsunami modeling and the PALP GPS are available on Zenodo:\\
\texttt{https://doi.org/10.5281/zenodo.5018003}


\begin{thebibliography}{67}
\expandafter\ifx\csname natexlab\endcsname\relax\def\natexlab#1{#1}\fi

\bibitem[Albin \& Bruno(2011)]{albinbruno}
Albin, N. \& Bruno, O.~P., 2011.
\newblock {A spectral FC solver for the compressible Navier--Stokes equations
  in general domains I: Explicit time-stepping}, {\it Journal of Computational
  Physics\/}, {\bf 230}(16), 6248--6270.

\bibitem[Altinok et~al.(2001)Altinok, Tinti, Alpar, Yalciner, Ersoy,
  Bortolucci, \& Armigliato]{altinok2001}
Altinok, Y., Tinti, S., Alpar, B., Yalciner, A., Ersoy, {\c{S}}., Bortolucci,
  E., \& Armigliato, A., 2001.
\newblock {The tsunami of August 17, 1999 in Izmit bay, Turkey}, {\it Natural
  Hazards\/}, {\bf 24}(2), 133--146.

\bibitem[Ambraseys et~al.(1994)Ambraseys, Melville, \& Adams]{ambraseys1994}
Ambraseys, N.~N., Melville, C.~P., \& Adams, R.~D., 1994.
\newblock {\it The Seismicity of Egypt, Arabia and the Red Sea\/}, Cambridge
  University Press.

\bibitem[Amlani \& Bruno(2016)]{amlanibruno}
Amlani, F. \& Bruno, O.~P., 2016.
\newblock {An FC-based spectral solver for elastodynamic problems in general
  three-dimensional domains}, {\it Journal of Computational Physics\/}, {\bf
  307}, 333--354.

\bibitem[Amlani \& Pahlevan(2020)]{amlaniniema}
Amlani, F. \& Pahlevan, N.~M., 2020.
\newblock {A stable high-order FC-based methodology for hemodynamic wave
  propagation}, {\it Journal of Computational Physics\/}, {\bf 405}, 109130.

\bibitem[Amlani et~al.(2019)Amlani, Bruno, L{\'o}pez-V{\'a}zquez, Trillo,
  Doval, Fern{\'a}ndez, \& Rodr{\'\i}guez-G{\'o}mez]{amlanicarlos}
Amlani, F., Bruno, O.~P., L{\'o}pez-V{\'a}zquez, J.~C., Trillo, C., Doval,
  {\'A}.~F., Fern{\'a}ndez, J.~L., \& Rodr{\'\i}guez-G{\'o}mez, P., 2019.
\newblock {Transient Propagation and Scattering of Quasi-Rayleigh Waves in
  Plates: Quantitative comparison between Pulsed TV-Holography Measurements and
  FC (Gram) elastodynamic simulations}, {\it arXiv preprint
  arXiv:1905.05289\/}.

\bibitem[Amlani et~al.(2021)Amlani, Wei, \& Pahlevan]{amlaniwei}
Amlani, F., Wei, H., \& Pahlevan, N.~M., 2021.
\newblock A new pseudo-spectral methodology without numerical diffusion for
  conducting dye simulations and particle residence time calculations, {\it
  arXiv preprint arXiv:2112.05257\/}.

\bibitem[Andrews(1976)]{andrews1976}
Andrews, D., 1976.
\newblock Rupture velocity of plane strain shear cracks, {\it Journal of
  Geophysical Research\/}, {\bf 81}(32), 5679--5687.

\bibitem[Archuleta(1984)]{archuleta1984}
Archuleta, R.~J., 1984.
\newblock {A faulting model for the 1979 Imperial Valley earthquake}, {\it
  Journal of Geophysical Research: Solid Earth\/}, {\bf 89}(B6), 4559--4585.

\bibitem[ASEAN(2018)]{asean2018}
ASEAN, 2018.
\newblock {Situation Update No. 15 - FINAL: M 7.4 Earthquake and Tsunami,
  Sulawesi, Indonesia}.

\bibitem[Bao et~al.(2019)Bao, Ampuero, Meng, Fielding, Liang, Milliner, Feng,
  \& Huang]{bao2019}
Bao, H., Ampuero, J.-P., Meng, L., Fielding, E.~J., Liang, C., Milliner, C.~W.,
  Feng, T., \& Huang, H., 2019.
\newblock {Early and persistent supershear rupture of the 2018 magnitude 7.5
  Palu earthquake}, {\it Nature Geoscience\/}, {\bf 12}(3), 200--205.

\bibitem[Bernard \& Baumont(2005)]{bernard2005}
Bernard, P. \& Baumont, D., 2005.
\newblock Shear mach wave characterization for kinematic fault rupture models
  with constant supershear rupture velocity, {\it Geophys. J. Int.\/}, {\bf
  162}, 431--447.

\bibitem[Bertiger et~al.(2010)Bertiger, Desai, Haines, Harvey, Moore, Owen, \&
  Weiss]{bertiger2010}
Bertiger, W., Desai, S.~D., Haines, B., Harvey, N., Moore, A.~W., Owen, S., \&
  Weiss, J.~P., 2010.
\newblock {Single receiver phase ambiguity resolution with GPS data}, {\it
  Journal of Geodesy\/}, {\bf 84}(5), 327--337.

\bibitem[Bouchon et~al.(2001)Bouchon, Bouin, Karabulut, Toks{\"o}z, Dietrich,
  \& Rosakis]{bouchon2001}
Bouchon, M., Bouin, M.-P., Karabulut, H., Toks{\"o}z, M.~N., Dietrich, M., \&
  Rosakis, A.~J., 2001.
\newblock {How fast is rupture during an earthquake? New insights from the 1999
  Turkey earthquakes}, {\it Geophysical Research Letters\/}, {\bf 28}(14),
  2723--2726.

\bibitem[Bruno \& Prieto(2014)]{brunoprieto}
Bruno, O.~P. \& Prieto, A., 2014.
\newblock {Spatially dispersionless, unconditionally stable FC--AD solvers for
  variable-coefficient PDEs}, {\it Journal of Scientific Computing\/}, {\bf
  58}(2), 331--366.

\bibitem[Bruno et~al.(2019)Bruno, Cubillos, \& Jimenez]{cubillos}
Bruno, O.~P., Cubillos, M., \& Jimenez, E., 2019.
\newblock {Higher-order implicit-explicit multi-domain compressible
  Navier-Stokes solvers}, {\it Journal of Computational Physics\/}, {\bf 391},
  322--346.

\bibitem[Bryant(2008)]{bryant2008tsunami}
Bryant, E., 2008.
\newblock {\it {Tsunami: The Underrated Hazard}\/}, Springer-Verlag, Berlin
  Heidelberg.

\bibitem[Burridge(1973)]{burridge1973}
Burridge, R., 1973.
\newblock Admissible speeds for plane-strain self-similar shear cracks with
  friction but lacking cohesion, {\it Geophysical Journal International\/},
  {\bf 35}(4), 439--455.

\bibitem[Carvajal et~al.(2019)Carvajal, Araya-Cornejo, Sep{\'u}lveda, Melnick,
  \& Haase]{carvajal2019}
Carvajal, M., Araya-Cornejo, C., Sep{\'u}lveda, I., Melnick, D., \& Haase,
  J.~S., 2019.
\newblock {Nearly instantaneous tsunamis following the Mw 7.5 2018 Palu
  earthquake}, {\it Geophysical Research Letters\/}, {\bf 46}(10), 5117--5126.

\bibitem[Dalguer \& Day(2007)]{dalguer2007}
Dalguer, L.~A. \& Day, S.~M., 2007.
\newblock Staggered-grid split-node method for spontaneous rupture simulation,
  {\it Journal of Geophysical Research: Solid Earth\/}, {\bf 112}(B2).

\bibitem[Das \& Aki(1977)]{das1977}
Das, S. \& Aki, K., 1977.
\newblock A numerical study of two-dimensional spontaneous rupture propagation,
  {\it Geophysical journal international\/}, {\bf 50}(3), 643--668.

\bibitem[Dunham \& Archuleta(2005)]{dunham2005}
Dunham, E.~M. \& Archuleta, R.~J., 2005.
\newblock Near-source ground motion from steady state dynamic rupture pulses,
  {\it Geophysical Research Letters\/}, {\bf 32}(3).

\bibitem[Dunham \& Bhat(2008)]{dunham2008a}
Dunham, E.~M. \& Bhat, H.~S., 2008.
\newblock Attenuation of radiated ground motion and stresses from
  three-dimensional supershear ruptures, {\it Journal of Geophysical Research:
  Solid Earth\/}, {\bf 113}(B8).

\bibitem[Dutykh \& Clamond(2016)]{dutykh2016}
Dutykh, D. \& Clamond, D., 2016.
\newblock Modified shallow water equations for significantly varying seabeds,
  {\it Applied mathematical modelling\/}, {\bf 40}(23-24), 9767--9787.

\bibitem[Elbanna et~al.(2021)Elbanna, Abdelmeguid, Ma, Amlani, Bhat, Synolakis,
  \& Rosakis]{elbanna2020}
Elbanna, A., Abdelmeguid, M., Ma, X., Amlani, F., Bhat, H.~S., Synolakis, C.,
  \& Rosakis, A.~J., 2021.
\newblock Anatomy of strike-slip fault tsunami genesis, {\it Proceedings of the
  National Academy of Sciences\/}, {\bf 118}(19).

\bibitem[Ellsworth et~al.(2004)Ellsworth, Celebi, Evans, Jensen, Kayen, Metz,
  Nyman, Roddick, Spudich, \& Stephens]{ellsworth2004b}
Ellsworth, W., Celebi, M., Evans, J., Jensen, E., Kayen, R., Metz, M., Nyman,
  D., Roddick, J., Spudich, P., \& Stephens, C., 2004.
\newblock {Near-field ground motion of the 2002 Denali fault, Alaska,
  earthquake recorded at pump station 10}, {\it Earthquake spectra\/}, {\bf
  20}(3), 597--615.

\bibitem[Favreau et~al.(2002)Favreau, Campillo, \& Ionescu]{favreau2002}
Favreau, P., Campillo, M., \& Ionescu, I.~R., 2002.
\newblock Initiation of shear instability in three-dimensional elastodynamics,
  {\it Journal of Geophysical Research: Solid Earth\/}, {\bf 107}(B7), ESE--4.

\bibitem[Fontana et~al.(2020)Fontana, Bruno, Mininni, \& Dmitruk]{fontanabruno}
Fontana, M., Bruno, O.~P., Mininni, P.~D., \& Dmitruk, P., 2020.
\newblock Fourier continuation method for incompressible fluids with
  boundaries, {\it Computer Physics Communications\/}, {\bf 256}, 107482.

\bibitem[Fritz et~al.(2018)Fritz, Synolakis, Kalligeris, Skanavis, Santoso,
  Rizal, Prasetya, Liu, \& Liu]{fritz2018}
Fritz, H.~M., Synolakis, C., Kalligeris, N., Skanavis, V., Santoso, F., Rizal,
  M., Prasetya, G.~S., Liu, Y., \& Liu, P.~L., 2018.
\newblock Field survey of the 28 {September 2018} {Sulawesi} tsunami, in {\em
  AGU Fall Meeting Abstracts\/}, vol. 2018, pp. NH22B--04.

\bibitem[Gaggioli et~al.(2019)Gaggioli, Bruno, \& Mitnik]{gaggiolibruno}
Gaggioli, E.~L., Bruno, O.~P., \& Mitnik, D.~M., 2019.
\newblock {Light transport with the equation of radiative transfer: The Fourier
  Continuation--Discrete Ordinates (FC--DOM) Method}, {\it Journal of
  Quantitative Spectroscopy and Radiative Transfer\/}, {\bf 236}, 106589.

\bibitem[He et~al.(2019)He, Feng, Li, Feng, Gao, \& Wu]{he2019}
He, L., Feng, G., Li, Z., Feng, Z., Gao, H., \& Wu, X., 2019.
\newblock {Source parameters and slip distribution of the 2018 Mw 7.5 Palu,
  Indonesia earthquake estimated from space-based geodesy}, {\it
  Tectonophysics\/}, {\bf 772}, 228216.

\bibitem[Heidarzadeh et~al.(2019)Heidarzadeh, Muhari, \&
  Wijanarto]{heidarzadeh2019}
Heidarzadeh, M., Muhari, A., \& Wijanarto, A.~B., 2019.
\newblock {Insights on the source of the 28 September 2018 Sulawesi tsunami,
  Indonesia based on spectral analyses and numerical simulations}, {\it Pure
  and Applied Geophysics\/}, {\bf 176}(1), 25--43.

\bibitem[Jamelot et~al.(2019)Jamelot, Gailler, Heinrich, Vallage, \&
  Champenois]{jamelot2019}
Jamelot, A., Gailler, A., Heinrich, P., Vallage, A., \& Champenois, J., 2019.
\newblock {Tsunami simulations of the Sulawesi Mw 7.5 event: Comparison of
  seismic sources issued from a tsunami warning context versus post-event
  finite source}, {\it Pure and Applied Geophysics\/}, {\bf 176}(8),
  3351--3376.

\bibitem[Johnson \& Beeson(2019)]{johnson2019}
Johnson, S.~Y. \& Beeson, J.~W., 2019.
\newblock {Shallow Structure and Geomorphology along the Offshore Northern San
  Andreas Fault, Tomales Point to Fort Ross, California}, {\it Bulletin of the
  Seismological Society of America\/}, {\bf 109}(3), 833--854.

\bibitem[Kajiura(1963)]{kajiura1963}
Kajiura, K., 1963.
\newblock The leading wave of a tsunami, {\it Bulletin of the Earthquake
  Research Institute, University of Tokyo\/}, {\bf 41}(3), 535--571.

\bibitem[Lotto et~al.(2017)Lotto, Nava, \& Dunham]{lotto2017}
Lotto, G.~C., Nava, G., \& Dunham, E.~M., 2017.
\newblock Should tsunami simulations include a nonzero initial horizontal
  velocity?, {\it Earth, Planets and Space\/}, {\bf 69}(1), 1--14.

\bibitem[Lyon \& Bruno(2010)]{lyonbrunoII}
Lyon, M. \& Bruno, O.~P., 2010.
\newblock {High-order unconditionally stable FC-AD solvers for general smooth
  domains II. Elliptic, parabolic and hyperbolic PDEs; theoretical
  considerations}, {\it Journal of Computational Physics\/}, {\bf 229}(9),
  3358--3381.

\bibitem[Mai(2019)]{mai2019}
Mai, P.~M., 2019.
\newblock Supershear tsunami disaster, {\it Nature Geoscience\/}, {\bf 12}(3),
  150--151.

\bibitem[Mello et~al.(2014)Mello, Bhat, Rosakis, \& Kanamori]{mello2014}
Mello, M., Bhat, H., Rosakis, A., \& Kanamori, H., 2014.
\newblock {Reproducing the supershear portion of the 2002 Denali earthquake
  rupture in laboratory}, {\it Earth and Planetary Science Letters\/}, {\bf
  387}, 89--96.

\bibitem[Muhari et~al.(2018)Muhari, Imamura, Arikawa, Hakim, \&
  Afriyanto]{muhari2018}
Muhari, A., Imamura, F., Arikawa, T., Hakim, A.~R., \& Afriyanto, B., 2018.
\newblock {Solving the puzzle of the September 2018 Palu, Indonesia, tsunami
  mystery: clues from the tsunami waveform and the initial field survey data},
  {\it Journal of Disaster Research\/}, {\bf 13}(Scientific Communication),
  sc20181108.

\bibitem[Oral et~al.(2020)Oral, Weng, \& Ampuero]{oral2020}
Oral, E., Weng, H., \& Ampuero, J.~P., 2020.
\newblock {Does a damaged-fault zone mitigate the near-field impact of
  supershear earthquakes? Application to the 2018 7.5 Palu, Indonesia,
  earthquake}, {\it Geophysical Research Letters\/}, {\bf 47}(1),
  e2019GL085649.

\bibitem[Passel{\`e}gue et~al.(2013)Passel{\`e}gue, Schubnel, Nielsen, Bhat, \&
  Madariaga]{passelegue2013}
Passel{\`e}gue, F.~X., Schubnel, A., Nielsen, S., Bhat, H.~S., \& Madariaga,
  R., 2013.
\newblock {From sub-Rayleigh to supershear ruptures during stick-slip
  experiments on crustal rocks}, {\it Science\/}, {\bf 340}(6137), 1208--1211.

\bibitem[Pedlosky(2013)]{pedlosky2013}
Pedlosky, J., 2013.
\newblock {\it Geophysical fluid dynamics\/}, vol. 710, Springer-Verlag, New
  York.

\bibitem[Pugh \& Woodworth(2014)]{pugh2014}
Pugh, D. \& Woodworth, P., 2014.
\newblock {\it Sea-level science: understanding tides, surges, tsunamis and
  mean sea-level changes\/}, Cambridge University Press.

\bibitem[Rebischung \& Schmid(2016)]{rebischung2016}
Rebischung, P. \& Schmid, R., 2016.
\newblock {IGS14/igs14. atx: a new framework for the IGS products}, in {\em AGU
  Fall Meeting 2016\/}.

\bibitem[Robinson et~al.(2006)Robinson, Brough, \& Das]{robinson2006a}
Robinson, D., Brough, C., \& Das, S., 2006.
\newblock {The Mw 7.8, 2001 Kunlunshan earthquake: Extreme rupture speed
  variability and effect of fault geometry}, {\it Journal of Geophysical
  Research: Solid Earth\/}, {\bf 111}(B8).

\bibitem[Robinson et~al.(2010)Robinson, Das, \& Searle]{robinson2010}
Robinson, D., Das, S., \& Searle, M., 2010.
\newblock Earthquake fault superhighways, {\it Tectonophysics\/}, {\bf
  493}(3-4), 236--243.

\bibitem[R{\"o}bke \& V{\"o}tt(2017)]{robke2017}
R{\"o}bke, B. \& V{\"o}tt, A., 2017.
\newblock The tsunami phenomenon, {\it Progress in Oceanography\/}, {\bf 159},
  296--322.

\bibitem[Rosakis et~al.(1999)Rosakis, Samudrala, \& Coker]{rosakis1999a}
Rosakis, A., Samudrala, O., \& Coker, D., 1999.
\newblock Cracks faster than the shear wave speed, {\it Science\/}, {\bf
  284}(5418), 1337--1340.

\bibitem[Sassa \& Takagawa(2019)]{sassa2019}
Sassa, S. \& Takagawa, T., 2019.
\newblock {Liquefied gravity flow-induced tsunami: first evidence and
  comparison from the 2018 Indonesia Sulawesi earthquake and tsunami
  disasters}, {\it Landslides\/}, {\bf 16}(1), 195--200.

\bibitem[Satake et~al.(2013)Satake, Fujii, Harada, \& Namegaya]{satake2013}
Satake, K., Fujii, Y., Harada, T., \& Namegaya, Y., 2013.
\newblock Time and space distribution of coseismic slip of the 2011 tohoku
  earthquake as inferred from tsunami waveform data, {\it Bulletin of the
  Seismological Society of America\/}, {\bf 103}(2B), 1473--1492.

\bibitem[Sep{\'u}lveda et~al.(2020)Sep{\'u}lveda, Haase, Carvajal, Xu, \&
  Liu]{sepulveda2020}
Sep{\'u}lveda, I., Haase, J.~S., Carvajal, M., Xu, X., \& Liu, P.~L., 2020.
\newblock {Modeling the sources of the 2018 Palu, Indonesia, tsunami using
  videos from social media}, {\it Journal of Geophysical Research: Solid
  Earth\/}, {\bf 125}(3), e2019JB018675.

\bibitem[Shahbazi et~al.(2013)Shahbazi, Hesthaven, \& Zhu]{shz}
Shahbazi, K., Hesthaven, J.~S., \& Zhu, X., 2013.
\newblock {Multi-dimensional hybrid Fourier continuation--WENO solvers for
  conservation laws}, {\it Journal of Computational Physics\/}, {\bf 253},
  209--225.

\bibitem[Socquet et~al.(2019)Socquet, Hollingsworth, Pathier, \&
  Bouchon]{socquet2019}
Socquet, A., Hollingsworth, J., Pathier, E., \& Bouchon, M., 2019.
\newblock {Evidence of supershear during the 2018 magnitude 7.5 Palu earthquake
  from space geodesy}, {\it Nature Geoscience\/}, {\bf 12}(3), 192--199.

\bibitem[Synolakis \& Okal(2005)]{synolakis2005}
Synolakis, C.~E. \& Okal, E.~A., 2005.
\newblock 1992--2002: perspective on a decade of post-tsunami surveys, in {\em
  Tsunamis: Case Studies and Recent Developments\/}, pp. 1--29, Springer,
  Dordrecht.

\bibitem[Tanioka \& Satake(1996)]{tanioka1996}
Tanioka, Y. \& Satake, K., 1996.
\newblock Tsunami generation by horizontal displacement of ocean bottom, {\it
  Geophysical research letters\/}, {\bf 23}(8), 861--864.

\bibitem[Tapponnier et~al.(1986)Tapponnier, Peltzer, \& Armijo]{tapponnier1986}
Tapponnier, P., Peltzer, G., \& Armijo, R., 1986.
\newblock On the mechanics of the collision between india and asia, {\it
  Geological Society, London, Special Publications\/}, {\bf 19}(1), 113--157.

\bibitem[Ulrich et~al.(2019)Ulrich, Vater, Madden, Behrens, van Dinther,
  Van~Zelst, Fielding, Liang, \& Gabriel]{ulrich2019}
Ulrich, T., Vater, S., Madden, E.~H., Behrens, J., van Dinther, Y., Van~Zelst,
  I., Fielding, E.~J., Liang, C., \& Gabriel, A.-A., 2019.
\newblock {Coupled, physics-based modeling reveals earthquake displacements are
  critical to the 2018 Palu, Sulawesi tsunami}, {\it Pure and Applied
  Geophysics\/}, {\bf 176}(10), 4069--4109.

\bibitem[Umar et~al.(2019)Umar, Margaglio, Fitrayansyah, et~al.]{syamsidik2019}
Umar, M., Margaglio, G., Fitrayansyah, A., et~al., 2019.
\newblock {Post-tsunami survey of the 28 September 2018 tsunami near Palu Bay
  in Central Sulawesi, Indonesia: Impacts and challenges to coastal
  communities}, {\it International Journal of Disaster Risk Reduction\/}, {\bf
  38}, 101229.

\bibitem[USGS(2018)]{usgs2018}
USGS, 2018.
\newblock {M 7.5 - 72 km N of Palu, Indonesia}.

\bibitem[Vigny(2003)]{vigny2003}
Vigny, C., 2003.
\newblock Present-day crustal deformation around sagaing fault, myanmar, {\it
  Journal of Geophysical Research\/}, {\bf 108}(B11).

\bibitem[Weatherall et~al.(2015)Weatherall, Marks, Jakobsson, Schmitt, Tani,
  Arndt, Rovere, Chayes, Ferrini, \& Wigley]{weatherall2015}
Weatherall, P., Marks, K.~M., Jakobsson, M., Schmitt, T., Tani, S., Arndt,
  J.~E., Rovere, M., Chayes, D., Ferrini, V., \& Wigley, R., 2015.
\newblock A new digital bathymetric model of the world{\textquotesingle}s
  oceans, {\it Earth and Space Science\/}, {\bf 2}(8), 331--345.

\bibitem[Webb(1997)]{webb1997}
Webb, F.~H., 1997.
\newblock {An Introduction to GIPsy/oasIs-II}, {\it JPL D-11088 (Jet Propulsion
  Laboratory)\/}.

\bibitem[Wu et~al.(1972)Wu, Thomson, \& Kuenzler]{wu1972}
Wu, F.~T., Thomson, K., \& Kuenzler, H., 1972.
\newblock Stick-slip propagation velocity and seismic source mechanism, {\it
  Bulletin of the Seismological Society of America\/}, {\bf 62}(6), 1621--1628.

\bibitem[Xia et~al.(2004)Xia, Rosakis, \& Kanamori]{xia2004}
Xia, K., Rosakis, A.~J., \& Kanamori, H., 2004.
\newblock {Laboratory earthquakes: The sub-Rayleigh-to-supershear rupture
  transition}, {\it Science\/}, {\bf 303}(5665), 1859--1861.

\bibitem[Yumul et~al.(2003)Yumul, Dimalanta, Tamayo, \& Maury]{yumul2003}
Yumul, G.~P., Dimalanta, C.~B., Tamayo, R.~A., \& Maury, R.~C., 2003.
\newblock Collision, subduction and accretion events in the philippines: A
  synthesis, {\it The Island Arc\/}, {\bf 12}(2), 77--91.

\bibitem[Zumberge et~al.(1997)Zumberge, Heflin, Jefferson, Watkins, \&
  Webb]{zumberge1997}
Zumberge, J., Heflin, M., Jefferson, D., Watkins, M., \& Webb, F., 1997.
\newblock {Precise point positioning for the efficient and robust analysis of
  GPS data from large networks}, {\it Journal of Geophysical Research: Solid
  Earth\/}, {\bf 102}(B3), 5005--5017.

\end{thebibliography}
\end{document}